\documentclass[letterpaper,a4paper] {IEEEtran}

\usepackage{graphicx}
\usepackage{epstopdf}
\usepackage{algpseudocode}

\usepackage{amsfonts,amssymb,enumerate}
\usepackage[centertags]{amsmath}
\usepackage{hyperref}
\usepackage{multirow}
\usepackage[utf8]{inputenc}
\usepackage{cite}

\newcommand{\ones}{\mathbf{1}}
\newcommand{\ud}{\mathop{}\!\mathrm{d}}
\newcommand{\reals}{\mathbb{R}}
\newcommand{\bx}{\mathbf{x}}
\newcommand{\bmu}{\boldsymbol{\mu}}
\newcommand{\cova}{\Sigma}
\newcommand{\cholcov}{\Lambda}
\newcommand{\var}{\mathbb{V}}

\newcommand{\bydef}{\stackrel{\Delta}{=}}

\newcommand{\bfy}{\mathbf{y}}
\newcommand{\bfx}{\mathbf{x}}
\newcommand{\bfu}{\mathbf{u}}
\newcommand{\vct}[1]{\mathbf{#1}}

\newcommand{\ie}{\textit{i}.\textit{e}.}

\newcommand{\y}{\mathbf{y}}
\newcommand{\x}{\mathbf{x}}
\newcommand{\z}{\mathbf{z}}
\newcommand{\bu}{\mathbf{u}}
\newcommand{\w}{\boldsymbol \omega}

\newcommand{\m}{\widehat{\x}}
\newcommand{\F}{\mathcal{F}}

\newcommand{\E}{\mathop{\mathbb E}}

\renewcommand{\t}[1]{\mathrm{T}#1}
\newcommand{\N}{\mathcal{N}}
\newcommand{\R}{\mathcal{R}}

\renewcommand{\d}[1]{\;\mathrm{d}#1}
\newcommand{\indi}{\mathbf{1}_{\R_i}(\x_t)}

\title{State Estimation for \\ Piecewise Affine State-Space Models}

\author{Rafael Rui, Tohid Ardeshiri, Henri Nurminen~\IEEEmembership{Student Member,~IEEE},\\ Alexandre Bazanella,~\IEEEmembership{Senior Member,~IEEE}, and Fredrik Gustafsson,~\IEEEmembership{Fellow,~IEEE}
\thanks{ R. Rui and A. Bazanella are  with Department of Electrical Engineering, Universidade Federal do Rio Grande do Sul, Porto Alegre 90040-060, Brazil (email: rafael.rui, bazanella@ufrgs.br) and are supported  by  Conselho Nacional de Desenvolvimento Científico e Tecnológico (CNPq). 
}
\thanks{T. Ardeshiri was with the Division of Automatic Control, Department of Electrical Engineering, Linköping University, Linköping, Sweden and received funding from Swedish research council (VR), project scalable Kalman filters for this work. 
T. Ardeshiri is currently with the Department of Engineering, University of Cambridge, Trumpington Street, Cambridge, CB2 1PZ, UK, (e-mail: ta417@cam.ac.uk).}\thanks{F. Gustafsson is with the Department of Electrical Engineering, Link\"{o}ping University, 58183 Link\"{o}ping, Sweden, (e-mails: fredrik@isy.liu.se).}\thanks{ H. Nurminen is  with the Department of Automation Science and Engineering, Tampere University of Technology (TUT), PO Box 692, 33101 Tampere, Finland  (e-mail: henri.nurminen@tut.fi). H. Nurminen receives funding from TUT Graduate School, the Foundation of Nokia Corporation, and Tekniikan edist\"amiss\"a\"ati\"o.
} 
}

\begin{document}
\maketitle

\begin{abstract}
We propose a filter for piecewise affine state-space (PWASS) models. In each filtering recursion, the true filtering posterior distribution is a mixture of truncated normal distributions. The proposed filter approximates the mixture with a single normal distribution via moment matching. The proposed algorithm is compared with the extended Kalman filter (EKF) in a numerical simulation
where the proposed method obtains, on average, better root mean square error (RMSE) than the EKF. 
\end{abstract}


%=============================================================================
%===============================  New SECTION  ===============================
%=============================================================================
\section{Introduction}
\label{sec:intro}

We consider a class of stochastic hybrid models in which the switch between submodels is not a jump Markov process, but it is state dependent. In hybrid models, the state domain can be divided into a number of regions, and within each region, the state dynamics are described by a set of differential equations. %The estimation problem in this kind of model consists of estimating the state hence the region the system is. 
Here we will deal with piecewise affine state-space (PWASS) models. PWASS models are a particular case of stochastic hybrid models, which are used to approximate nonlinear dynamical systems and have been considered in several fields, such as automatic control \cite{ChzeEngSeah2009a}, signal processing \cite{Doucet2001}, system biology \cite{Porreca2009}, and computer vision \cite{Vidal2006}. 
\begin{figure}[t]
\centering
\includegraphics[width=0.8\columnwidth]{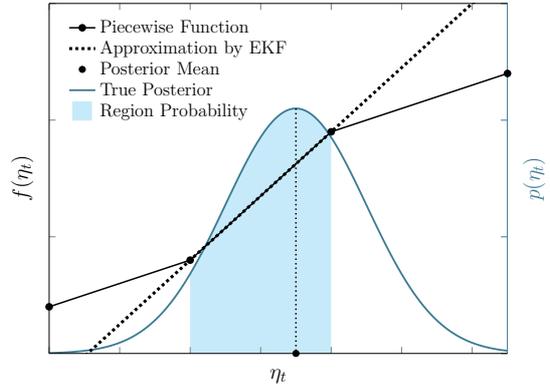}%
\caption{A piecewise affine function $f(\eta_t)$ is locally approximated by a single line in EKF, while the proposed filter computes the first two moments of the posterior over the entire support. }%
\label{fig:posterior}%
\end{figure}

 Most of studies in the literature on Bayesian filtering of stochastic hybrid systems are limited to jump Markov systems \cite{Murphy1998,Ghahramani2000,Barber2006a,Barber2007,Mesot2008,Ozkan2013} or the so called semi-Markov jump linear systems \cite{Hwang2006,Blom2007,ChzeEngSeah2009a,Capponi2010}.
However, in practice, there are systems where the jump Markov model for transitions between submodels is an approximation of the reality. For example, in the JAS 39 Gripen aircraft, the dynamic of the pitch rate in the model for the flight dynamic in the longitudinal direction has a nonlinear dependence on the angle of attack \cite{Larsson2012}. Further, this nonlinear dependence is modeled by a piecewise affine function. 

The exact Bayesian filtering solution for such systems is a mixture of truncated normal distributions, where the number of mixture components grows exponentially with time. When the extended Kalman filter (EKF) is used in PWASS models,  the piecewise affine function is approximated by a single line, as showed in Fig. \ref{fig:posterior}. This is problematic when the state uncertainty is large compared to the sizes of the regions. In this letter, we propose a Bayesian filtering algorithm for PWASS models that uses the exact time and measurement Kalman filter updates for each submodel avoiding linearization errors. In the proposed filter  the cumulative distribution function (CDF) is used to compute the posterior distribution of the state as well as the probability of each region (shaded area in Fig. \ref{fig:posterior}). The mixture explosion is avoided through approximating each posterior mixture of  truncated normal distributions by a single normal distribution with matched moments.

%=============================================================================
%===============================  New SECTION  ===============================
%=============================================================================
\section{Problem formulation}
Consider the PWASS model \cite{Andrea2012}
\begin{subequations}
\label{eq:nlss}
\begin{align}
\x_{t+1}=\ &\F(\x_t)+B \bu_t + \w_t,\\
\y_t=\ &C\x_t+\boldsymbol\nu_t,
\end{align}
\end{subequations}
where $\bfy_t\in \mathbb{R}^{n_y}$ is the measurement; $C \in \mathbb{R}^{n_y\times n_x}$ is the measurement matrix; $\bfu_t\in \mathbb{R}^{n_u}$ is the deterministic input; $B\in \mathbb{R}^{n_x \times n_u}$ is the input matrix;  $\boldsymbol \omega_t\in \mathbb{R}^{n_x}$ and $\boldsymbol \nu_t\in \mathbb{R}^{n_y}$ are the process and measurement noise terms respectively;  $\bfx_t\in \mathbb{R}^{n_{x}}$ is the state vector partitioned by two scalar variables $\eta_t$ and $\zeta_t$ as well as  a vector $\boldsymbol  \chi_t\in \mathbb{R}^{(n_{x}-2)}$ as in $\x_t\triangleq \begin{bmatrix}  \eta_t,\zeta_t,\boldsymbol  \chi_t^\t \end{bmatrix}^\t$.
%\begin{align}
%\x_t\triangleq \begin{bmatrix} \chi_t \\ \zeta_t \\ \eta_t \end{bmatrix},&&  ; \label{eq:statepart}
%\end{align}
The nonlinear function $\F(\cdot)$ is the state transition function with the following structure 
\begin{align}
\F(\x_t)\triangleq \begin{bmatrix}\Phi^\t \x_t \\ f(\eta_t) + \boldsymbol\phi^\t \z_t \\ F\x_t   \end{bmatrix}, \label{eq:pw_geral}
\end{align}
where $\z_t\triangleq \begin{bmatrix} \zeta_t,\boldsymbol{\chi}_t^\t \end{bmatrix}^\t$, $\boldsymbol \phi \in \mathbb{R}^{(n_x-1)}$, $\Phi \in \mathbb{R}^{n_x}$, $F \in \mathbb{R}^{(n_x-2) \times n_x}$; and  the piecewise affine function $f(\eta_t)$ is given by
\begin{equation}
f(\eta_t) = 
 \begin{cases}
 f_1 =  a_1 {\eta}_{t} + b_1 & \text{if } {l}_{1}<{\eta}_{t}\leq{l}_{2} \\
 %f_2 =  a_2 {\eta}_{t} + b_2 & \text{if } {l}_{2}<{\eta}_{t}\leq{l}_{3} \\
	\quad \vdots & \\
	   f_{N_r} = a_{N_{r}} {\eta}_{t} + b_{N_{r}} & \text{if } {l}_{{N_{r}}}<{\eta}_{t}<{l}_{{N_{r+1}}}, \\
  \end{cases}
\label{eq:PWregion}
\end{equation}	
with $l_1 = -\infty$ and $l_{N_r+1} = + \infty$. For a given region $\mathcal R_i \triangleq \left\{\x_t :  {l}_{i}<{\eta}_{t}\leq{l}_{i+1}\right\}$ it is possible to rewrite \eqref{eq:pw_geral} using \eqref{eq:PWregion} as 
\begin{align}
\F(\x_t) &=  \begin{bmatrix} 
\Phi^\t \x_t \\a_i {\eta}_{t} + b_i +  \boldsymbol\phi^\t \z_t  \\ F\x_t    
\end{bmatrix} \\
&= \overbrace{ 
\begin{bmatrix} 
\Phi^\t   \\ [a_i  ~ \boldsymbol \phi^\t] \\ F 
\end{bmatrix}}^{A_i}\x_t  + 
\overbrace{
\begin{bmatrix} 
\mathbf{0} \\ b_i  \\ \mathbf{0}   
\end{bmatrix}}^{\mathbf{b}_i} \label{eq:pw_geral2}\\  
&=  A_i\x_t + \mathbf{b}_i .
\label{eq:pw_geral23}
\end{align}
Hence, for a given  region $\mathcal R_i$, the model \eqref{eq:nlss} can be written as the conditionally affine state-space model
\begin{subequations}
\label{eq:nlss2}
\begin{align}
\x_{t+1}=\ &A_i \x_t+B \bu_t +\mathbf b_i + \w_t,\\
\y_t=\ &C\x_t+\boldsymbol\nu_t,
\end{align}
\end{subequations}
where $A_i$ and $\mathbf{b}_i$ are defined in \eqref{eq:pw_geral2}. The index $i \in \left\{1,\cdots, N_{r}\right\}$ determines in which piecewise affine dynamics the system is at time $t$, \ie, which submodel is active at time $t$. The initial state has a prior distribution $	\mathbf{\bfx}_1 \sim \mathcal{N}(\hat \bfx_{1|0},P_{1|0})$, where $\N(\boldsymbol{\mu}, \Sigma)$ denotes a Gaussian distribution with mean $\boldsymbol{\mu}$ and covariance $\Sigma$, and  the subscript ``$t_1|t_2$" is read ``at time $t_1$ using measurements up to time $t_2$". Also, we assume $\{\boldsymbol \omega_t\in\mathbb{R}^{n_x}| 1\leq t \leq T\}$  and $\{\boldsymbol \nu_t\in\mathbb{R}^{n_y}| 1\leq t \leq T\}$  are mutually independent white Gaussian noise sequences with covariance $Q$ and $R$ respectively.
In this letter, we propose a filter to estimate $p(\x_{t}|\y_{1:t})$.

%=============================================================================
%===============================  New SECTION  ===============================
%=============================================================================
\section{Proposed Solution}
Assume that at time $t$ the following filtering posterior distribution for $\x_t$  is available
\begin{align}
p(\x_t|\y_{1:t})=\N(\x_t;\m_{t|t},P_{t|t}).
\end{align}
This distribution can be rewritten using the indicator function as in
\begin{align}
p(\x_t|\y_{1:t})=\sum_{i=1}^{N_{r}} \indi \N(\x_t;\m_{t|t},P_{t|t}),
\end{align}
where
\begin{align}
\mathbf{1}_{\mathcal A}(\x) \triangleq
\begin{cases} 
1 &\text{if } \x \in \mathcal A, \\
0 &\text{if } \x \notin  \mathcal A.
\end{cases}
\label{eq:delta}
\end{align}
Using \eqref{eq:nlss2}, the state transition density $p(\x_{t+1}|\x_t)$ and the likelihood function $p(\y_{t+1}|\x_{t+1})$ can be written as
 \begin{align}
p(\x_{t+1}|\x_t) &= \N(\x_{t+1};A_i \x_t+B \bu_t +\mathbf b_i,Q),\label{eq:trans_dens}\\
p(\y_{t+1}|\x_{t+1}) &= \N(\y_{t+1};C\x_{t+1},R).
\label{eq:like}
\end{align}
 Therefore, the joint posterior $p(\x_{t},\x_{t+1},\y_{t+1}|\y_{1:t})$ can be written as
\begin{align}
&p(\x_{t},\x_{t+1},\y_{t+1}|\y_{1:t})=\sum_{i=1}^{N_{r}} \indi \N(\x_t;\m_{t|t},P_{t|t})\nonumber\\
& \times \N(\x_{t+1};A_i \x_t+B \bu_t +\mathbf b_i,Q)  \N(\y_{t+1};C\x_{t+1},R),
\end{align}
which can be rewritten in matrix form as
\begin{align}
p(&\x_{t},\x_{t+1},\y_{t+1}|\y_{1:t})\nonumber\\
&=\sum_{i=1}^{N_{r}} \indi \N([\x_{t}^\t,\x_{t+1}^\t,\y_{t+1}^\t]^\t;\boldsymbol{\mu}_{t,i},\Sigma_{t,i}), \label{eq:jointposterior}
\end{align}
where
\begin{align}
\boldsymbol{\mu}_{t,i}\bydef \left[
\begin{array}{c} 
\multirow{2}{*}{$\boldsymbol{\mu}_{1i}$} \\ \\ \hline
\boldsymbol{\mu}_{2i}
 \end{array}
\right] = 
&\left[
\begin{array}{c}
\m_{t|t}\\ 
 A_i \m_{t|t}+B \bu_t +\mathbf b_i \\ \hline
CA_i \m_{t|t}+C(B \bu_t +\mathbf b_i)
\end{array}
\right]
\end{align} 
and
\small
\begin{align}
&\Sigma_{t,i} \triangleq \left[\begin{array}{c|c}
\Sigma_{11i}& \Sigma_{12i}      \\ \hline
\Sigma_{21i}& \Sigma_{22i}
\end{array} 
\right]=\\
&\left[
\begin{array}{cc|c}
 P_{t|t}                        &   P_{t|t}A_i^\t                &    (P_{t|t}A_i^T)C^T    \\  
 A_i P_{t|t}                    &   A_i P_{t|t} A_i^\t+Q         &    (A_i P_{t|t} A_i^\t+Q)^TC^T       \\    \hline
 C(A_i P_{t|t})                 &   C(A_i P_{t|t} A_i^\t+Q)      &    C(A_i P_{t|t} A_i^\t+Q)C^\t+R               \\
\end{array}
\right].
\end{align}
\normalsize
The conditional distribution of $\x_{t}$ and $\x_{t+1}$ given $\y_{t+1}$ is 
%=============================================================================
\begin{align}
p&(\x_{t},\x_{t+1} | \y_{1:t+1}) = \frac{1}{Z_t}p(\x_{t},\x_{t+1},\y_{t+1}|\y_{1:t}) \\
 &=\frac{1}{Z_t}\sum_{i=1}^{N_{r}} \indi \N([\x_{t}^\t,\x_{t+1}^\t,\y_{t+1}^\t]^\t;\boldsymbol{\mu}_{t,i},\Sigma_{t,i}) \\
 &=\frac{1}{Z_t}\sum_{i=1}^{N_{r}} \indi \N(\y_{t+1}; \boldsymbol{\mu}_{2i},\Sigma_{22i})\nonumber \\
& \hspace{3cm} \times \N([\x_{t}^\t,\x_{t+1}^\t]^\t; \tilde{\boldsymbol{\mu}}_i,\tilde{\Sigma}_i),  \label{eq:posteriormixture}
\end{align}
where
\begin{align}
\tilde{\boldsymbol{\mu}}_i&\bydef \left[\begin{array}{c} \tilde{\boldsymbol{\mu}}_{1i}\\ \hline
\tilde{\boldsymbol{\mu}}_{2i} 
 \end{array}\right] = \boldsymbol{\mu}_{1i} + \Sigma_{21i}^T \Sigma_{22i}^{-1}(\y_{t+1} - \boldsymbol{\mu}_{2i}), \label{eq:mu_tilda_i}\\
\tilde{\Sigma}_i  &\bydef 
\left[\begin{array}{c|c}
\tilde{\Sigma}_{11i}& \tilde{\Sigma}_{12i}      \\ \hline
\tilde{\Sigma}_{21i}& \tilde{\Sigma}_{22i}
\end{array}
\right] = \Sigma_{11i} - \Sigma_{21i}^T\Sigma_{22i}^{-1} \Sigma_{21i},
\label{eq:sigma_tilda_i}
\end{align}
$Z_t$ is a normalizing constant, and the partitions in \eqref{eq:mu_tilda_i}  and \eqref{eq:sigma_tilda_i} have equal dimensions. 
%The posterior distribution of the regions can be computed using the proposed method.
 The quantities $\Pr(\x_t\in \R_i|\y_{1:t+1})$ for $i=1 \cdots N_{r}$ as well as the normalizing constant $Z_t$ can be computed  via integration of $p(\x_{t},\x_{t+1} | \y_{1:t+1})$ as in 
\begin{align}
&\Pr(\x_t\in \R_i|\y_{1:t+1})  \nonumber\\
&=\frac{1}{Z_t}\N(\y_{t+1}; \boldsymbol{\mu}_{2i},\Sigma_{22i}) \int_{\R_i} \N(\x_t; \tilde{\boldsymbol{\mu}}_{1i},\tilde{\Sigma}_{11i})  \d \x_t\\
&=\frac{1}{Z_t}\N(\y_{t+1}; \boldsymbol{\mu}_{2i},\Sigma_{22i}) \; \Gamma_i, 
\label{eq:integral}
\end{align}
where
\begin{align}
& \Gamma_i \bydef  \int_{ {l}_{i}<{\eta}_{t}\leq{l}_{i+1}} \N(\eta_t; [\tilde{\boldsymbol{\mu}}_{1i}]_{(1,1)},[\tilde{\Sigma}_{11i}]_{(1,1)})  \d \eta_t\\
&=\frac{1}{2}\text{erf}\left(\frac{l_{i+1}- [\tilde{\boldsymbol{\mu}}_{1i}]_{(1,1)}}{\sqrt{2[\tilde{\Sigma}_{11i}]_{(1,1)}}}\right) - \frac{1}{2}\text{erf}\left(\frac{l_{i}- [\tilde{\boldsymbol{\mu}}_{1i}]_{(1,1)}}{\sqrt{2[\tilde{\Sigma}_{11i}]_{(1,1)}}}\right),
\label{eq:integral_sol}
\end{align}
where the $\text{erf}(\cdot)$ is the error function, $[\Sigma]_{(i,j)}$ is the element in row $i$ and column $j$ of $\Sigma$, and 
\begin{equation}
{Z_t}=\sum_{i=1}^{N_r}N(\y_{t+1}; \boldsymbol{\mu}_{2i},\Sigma_{22i}) \;\Gamma_i.
\label{eq:Zt}
\end{equation}
The probability $\Pr(\x_t\in \R_i|\y_{1:t+1})$ represents the probability that the state be in the region $\R_i$ at time $t$ given all information up to time $t+1$. 
The filtering posterior distribution $p(\x_{t+1}|\y_{1:t+1})$ can be computed via the integration
\begin{align}
&p(\x_{t+1}|\y_{1:t+1})=\frac{1}{Z_t}\int  {p(\x_{t},\x_{t+1},\y_{t+1}|\y_{1:t}) \d \x_t }\\
&=\frac{1}{Z_t} \sum_{i=1}^{N_{r}} \N(\y_{t+1}; \boldsymbol{\mu}_{2i},\Sigma_{22i})    \nonumber\\
&\hspace{1cm} \times \int_{\R_i} \indi   \N([\x_{t}^\t,\x_{t+1}^\t]^\t; \tilde{\boldsymbol{\mu}}_i,\tilde{\Sigma}_i)   \d \x_t . \label{eq:marginalposteriormixture}
%&=\frac{1}{Z_t}    \sum_{i=1}^{N_{r}}  \N(\y_{t+1}; \boldsymbol{\mu}_{1i},\Sigma_{11i}) \\
%&\hspace{3cm}\int_{\R_i}   \N([\x_{t+1}^\t,\x_t^\t]^\t; \tilde{\boldsymbol{\mu}}_i,\tilde{\Sigma}_i)   \d \x_t  
\end{align}

The joint posterior distribution on the right-hand side of~\eqref{eq:marginalposteriormixture} is a mixture of doubly  truncated multivariate normal distributions (DTMND) \cite{Nurminen2016}. 
In order to have a recursive algorithm, the posterior will be approximated by a normal distribution. To this end, the mean and the covariance of the posterior distribution $p(\x_{t+1}|\y_{1:t+1})$ is needed. The mean and covariance of a mixture distribution can be computed using the mean and the covariance of the components of the mixture density via standard moment matching formulas \cite{Runnalls2007}. Hence, the problem boils down to computing the mean and the covariance of the DTMND which is presented in the Apendix \ref{ap1}. 
 
The proposed filter for PWASS models will be referred to by PAKF (Piecewise Affine Kalman Filter). The filtering recursion is given in TABLE~\ref{table:filtering}. The expressions for computing the mean and the covariance of a DTMND for a given region $\mathcal R_i$ are given in the lines $18$ and $19$ of TABLE \ref{table:filtering}. The probabilities $\Pr(\x_t\in \R_i|\y_{1:t+1})$ as well as the normalizing constant $Z_t$ are calculated in the lines 21 and 22 and are used within the moment matching whose formulas are given in the lines $23$ and $24$ of TABLE \ref{table:filtering}.
%The complete filtering algorithm for computing these moments as well as the parameters of the posterior~\eqref{eq:marginalposteriormixture} are given in TABLE~\ref{table:filtering}. The proposed filter will be referred to by PAKF (Piecewise affine Kalman Filter) in the rest of this letter.
%===========================================================================

%=============================================================================
\begin{table}[t]
\caption{Filtering recursion for PWASS models using the Piecewise Affine Kalman Filter (PAKF)}\label{table:filtering}
\vspace{-1mm}\rule{\columnwidth}{1pt}%\vspace{0mm}
\begin{algorithmic}[1]
%------------------------------
\State \textbf{Inputs:} $A_i, \mathbf b_i, l_i,  ~ i=1\dots, N_r$, $B$, $C$, $Q$, $R$, $\mathbf u_{t}$, $\y_{t+1},\m_{t|t},P_{t|t}$
	\For{$i$ = 1 to $N_r$}
	\Statex \textit{Kalman filter prediction step}
		\State {
		$\boldsymbol{\mu}_{1i} \gets  
			\left[
				\begin{smallmatrix}
				 	\m_{t|t}\\
					A_i \m_{t|t}+B \bu_t +\mathbf b_i 
				\end{smallmatrix}
			\right]$}
		\State { $\boldsymbol{\mu}_{2i} \gets CA_i \m_{t|t}+C(B \bu_t +\mathbf b_i)$} 
		\State {  $ \Sigma_{11i}\gets \left[
\begin{smallmatrix}
    P_{t|t}                 &     P_{t|t}A_i^\t   \\
    A_i P_{t|t}             &     A_i P_{t|t} A_i^\t+Q    \\
\end{smallmatrix}
\right] $}
		\State {  $ \Sigma_{22i} \gets C(A_i P_{t|t} A_i^\t+Q)C^\t+R  $}
		\State { $ \Sigma_{21i} \gets \left[
		\begin{smallmatrix}
			 C(A_i P_{t|t})       &     C(A_i P_{t|t} A_i^\t+Q)
	   \end{smallmatrix}	\right] $}
		\Statex \textit{Kalman filter update step}
		\State {  $ 
			\tilde{\boldsymbol{\mu}}_i  \gets \boldsymbol{\mu}_{1i} + \Sigma_{21i}^T \Sigma_{22i}^{-1}(y_{t+1} - \boldsymbol{\mu}_{2i}) $}
		\State { $ \tilde{\Sigma}_i \gets  \Sigma_{11i} - \Sigma_{21i}^T\Sigma_{22i}^{-1} \Sigma_{21i}  $}
		\Statex \textit{Computing integral \eqref{eq:integral}}
		\State { $ w_i \gets   \frac{1}{2}\text{erf}\left(\frac{l_{i+1}- [\tilde{\boldsymbol{\mu}}_{1i}]_{(1,1)}}{\sqrt{2[\tilde{\Sigma}_{11i}]_{(1,1)}}}\right) - \frac{1}{2}\text{erf}\left(\frac{l_{i}- [\tilde{\boldsymbol{\mu}}_{1i}]_{(1,1)}}{\sqrt{2[\tilde{\Sigma}_{11i}]_{(1,1)}}}\right) $ }
		
	\State { $ w_i \gets  	\N(\y_{t+1}; \boldsymbol{\mu}_{2i},\Sigma_{22i}) \times w_i$ }
	
		\Statex \textit{Computing mean and covariance of the DTMND}
		%\State $\Lambda \gets \mathrm{chol}(\tilde{\Sigma}_i,\verb+'lower'+)$ \Comment {(\textsf{Matlab} syntax)}
		\State $\Lambda \gets \mathrm{chol}(\tilde{\Sigma}_i,'lower')$ \Comment {(Cholesky decomposition)}
		\State $\lambda_1 \gets \frac{l_i-[\tilde{\boldsymbol{\mu}}_i]_{(1,1)} }{[\Lambda ]_{(1,1)}}$
		\State $\lambda_2 \gets \frac{l_{i+1}-[\tilde{\boldsymbol{\mu}}_i]_{(1,1)} }{[\Lambda ]_{(1,1)} }$
		\State $\mathcal Z \gets \frac{1}{2}\text{erf}\left(\frac{\lambda_2}{\sqrt{2}}\right)-\frac{1}{2}\text{erf}\left(\frac{\lambda_1}{\sqrt{2}}\right)$
		\State $m_{i1} \gets \frac{\N(\lambda_1;0,1)-\N(\lambda_2;0,1)}{\mathcal Z}$
		\State $s_{i1} \gets 1+\frac{\lambda_1 \N(\lambda_1;0,1)-\lambda_2\N(\lambda_2;0,1)}{\mathcal Z} - ({m}_{i1})^2$
		\State $\mathbf{m}_i \gets \Lambda \left[\begin{smallmatrix}  m_{i1}  \\ \mathbf{0}_{2n_x-1} \end{smallmatrix}\right] + \tilde{\boldsymbol{\mu}}_i$
		\State $S_i \gets \Lambda \left[\begin{smallmatrix} s_{i1}  & \mathbf{0}_{2n_x-1}^\t \\ \mathbf{0}_{2n_x-1} & I_{2n_x-1}  \end{smallmatrix}\right] \Lambda^\t$
	\EndFor
 \Statex \textit{Normalizing constant \eqref{eq:Zt}}
	\State $Z_t \gets {\sum_{i=1}^{N_r}w_i}$ 
	\State $w_i \gets \frac{w_i}{Z_t}$ 
	\Statex \textit{Moment matching}
	\State $\widehat{\mathbf{m}}_{t} \gets \sum_{i=1}^{N_r} w_i ~\mathbf{m}_i $

		\State $ P_{t} \gets \sum_{i=1}^{N_r} w_i\left(S_i +  (\mathbf{m}_i - \widehat{\mathbf{m}}_{t})(\mathbf{m}_i - \widehat{\mathbf{m}}_{t})^\t\right)$

	\State $\m_{t+1|t+1} \gets [\widehat{\mathbf{m}}_{t}]_{n_x+1:2n_x} $
	\State $P_{t+1|t+1} \gets [P_{t}]_{n_x+1:2n_x,n_x+1:2n_x}$
\State \textbf{Outputs: $\m_{t+1|t+1}$ and  $P_{t+1|t+1}$ } 

%------------------------------
\end{algorithmic}
\noindent \rule{\columnwidth}{1pt}\vspace{0mm}
\end{table}
%=============================================================================
%\clearpage
%=============================================================================
%===============================  New SECTION  ===============================
%=============================================================================
\section{Numerical Simulations}

Numerical simulations are performed to evaluate the performance of PAKF. In these simulations, PAKF is compared to the extended Kalman filter (EKF) and the marginalized particle filter (MPF) \cite{Doucet2000a,Schon2005}.  The EKF expressions for PWASS models are those in the lines $3\text{-}9$ of TABLE \ref{table:filtering}. In EKF, they are evaluated only for the region where $\m_{t|t}$ is located at time $t$. The MPF is used to compute the optimal Bayesian solution. This optimal solution  will be used as a reference in the evaluation of PAKF. 
%All numerical computations are done using MATLAB in a Core2Quad 2.66GHz computer.  
All numerical computations are done using MATLAB.
\begin{table}[t]
\centering
\caption{Parameters of the SDOFS model}
\label{table:params}
\vspace{-.1cm}
\begin{tabular}{cccccc}
Param.&$\Delta t$ (s) & $D$ (N$\cdot$s/mm)&  $M$ (t) & $a$ (N/mm) & $l_1,l_2$ (mm)  \\
\hline
\rule{0pt}{2ex}
 Value &0.01 & 1 & 1 & $ [50 ~  5 ~  50]$ & $1$  \\
\end{tabular}
\vspace{-6mm}
\end{table} 
 
\begin{figure}[t]%
\centering
\includegraphics[width=0.6\columnwidth]{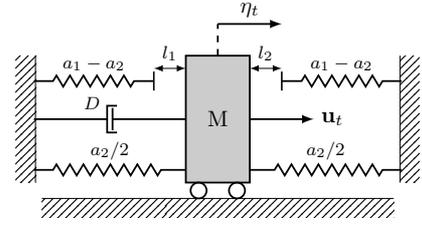}%
\caption{SDOFS physical model with piecewise affine spring characteristics.}%
\label{fig:mass-spring}%
\end{figure}

\begin{figure}%
\centering
\includegraphics[width=0.5\columnwidth]{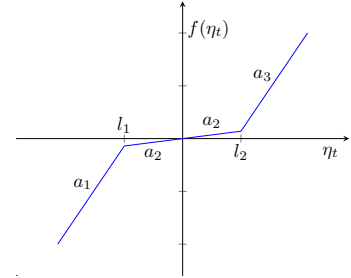}%
\caption{Piecewise affine spring characteristics of model in Fig.~\ref{fig:mass-spring} .}%
\label{fig:spring_pw}%
\end{figure}

Nonlinear vibrations caused by clearance can be modeled as a single-degree-of-freedom system (SDOFS) with piecewise affine spring characteristics \cite{Mahfouz1990}. Fig.~\ref{fig:mass-spring} shows the physical model of SDOFS and Fig. \ref{fig:spring_pw} presents its piecewise affine spring characteristic. The discretized PWASS model for the  SDOFS can be written as 
\begin{subequations}
\begin{align}
\x_{t+1}&\bydef \begin{bmatrix}
	\eta_{t+1} \\ \zeta_{t+1}
\end{bmatrix} = \begin{bmatrix}
	1 & \Delta t\\ -\frac{\Delta ta_i}{M} & 1- \frac{\Delta t D}{M}
\end{bmatrix}\x_t \nonumber \\ 
 & \hspace{1.7cm}  + \begin{bmatrix}
	0 \\ \frac{\Delta t}{M}
\end{bmatrix} \bfu_t + \begin{bmatrix}
	0 \\ -\frac{\Delta t b_i}{M}
\end{bmatrix} + \w_t,\\
\y_t &= \begin{bmatrix}
	1&0
\end{bmatrix}\x_{t} + \boldsymbol\nu_t ,
\label{eq:modely}%
\end{align}
\label{eq:model}%
\end{subequations}
where $\eta_{t}$ and $\zeta_{t}$ are the position in [mm] and the velocity in [mm/s] of the mass $M$, respectively, $\Delta t$ is the sampling time, $D$ is the damping coefficient, and $a_i$ and $b_i$ are the piecewise affine spring coefficients such that
\begin{equation}
f(\eta_t) = 
 \begin{cases}
 f_1 =  a_1 {\eta}_{t} + b_1 & \text{if } -\infty<{\eta}_{t}\leq{l}_{1} \\
 f_2 =  a_2 {\eta}_{t} + b_2 & \text{if } \quad ~ {l}_{1}<{\eta}_{t}\leq{l}_{2} \\
 f_3 =  a_3 {\eta}_{t} + b_3 & \text{if } \quad ~ l_2<{\eta}_{t}<+\infty 
  \end{cases}
\label{eq:PWregion_ex}
\end{equation}	
with $a_3 = a_1, b_1 = l_1 (a_2 - a_1), b_2 = 0$ and  $b_3 = l_2 (a_2 - a_3)$. 
Monte Carlo (MC) simulations are performed, where  the model \eqref{eq:model} is simulated for $T=400$ time steps. The parameters values are given in TABLE~\ref{table:params}. Further, $\mathbf{\bfx}_1$ is sampled from standard bivariate normal distribution, $\boldsymbol \nu_t \sim \mathcal{N}(0,1)$,  $\w_t \sim \mathcal{N}([0 ~ 0]^T,\text{diag} [0.01 \ 0.01])$  and $\bfu_t  \sim \mathcal{N}(0,5^2)$. For the MPF, we use 10\:000 particles. 

\begin{figure}[ht]%
\centering
%\hspace{0.4cm}
\includegraphics[width=0.75\columnwidth]{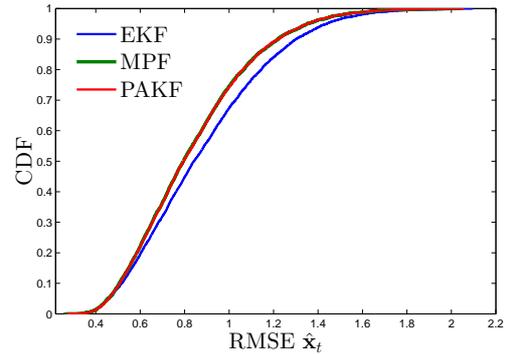}%
\caption{Cumulative distribution of the RMSE for EKF, MPF and PAKF.}%
\label{fig:hist} %
\end{figure}

We compare the three filters in terms of the root mean square error (RMSE) between the true state and the predicted state
\small
\begin{equation}
\text{RMSE}({\x}^{(j)}) = \sqrt{\frac{1}{2T} \sum_{t=1}^T\left( (\eta_t^{(j)} - \hat\eta_{t|t}^{(j)})^2  + (\zeta_t^{(j)} - \hat\zeta_{t|t}^{(j)})^2\right)},
\label{eq:RMSE}
\end{equation}
\normalsize
where  $\m_{t|t}^{(j)} = \left[\hat\eta_{t|t}^{(j)} ~ \hat\zeta_{t|t}^{(j)}\right]^\t$ and $\x_t^{(j)}= \left[\eta_t^{(j)} ~ \zeta_t^{(j)}\right]^\t $   denote the estimated mean of the state $\x_t$ and its true value in the $j$th MC run, respectively.
 Columns two and three of Table \ref{table:statistic}  show the average over 5\:000 MC simulations of RMSE (ARMSE) for each filter as well as the standard deviation of the RMSE (STD). We noticed that the ARMSE for PAKF is $5.35\%$ smaller than that of the EKF. The ARMSE for MPF is $0.12\%$ smaller than that of the PAKF. We also noticed that EKF has the highest STD of all the filters. The Fig. \ref{fig:hist} presents the cumulative distribution of the RMSE for each filter. The minimum and maximum RMSE values for each filter are presented in the last two columns of TABLE \ref{table:statistic}. 
\begin{table}[t]
\centering
\caption{Filter comparison results.}
\vspace{-.1cm}
\label{table:statistic}
\begin{tabular}{c|cc|cc}
Filter & ARMSE & STD & min. RMSE & max. RMSE \\ \hline
EKF    &         0.88327 & 0.30612 &  0.25792  &  2.09761                     \\
PAKF   &         0.83600 & 0.27931 &  0.27256 & 2.05709                               \\
MPF     &        0.83505 & 0.27994  & 0.26467 & 2.04196 	                        
\end{tabular}
\end{table}

\begin{figure}
\centering
%\hspace{0.3cm}
\includegraphics[width=0.75\columnwidth]{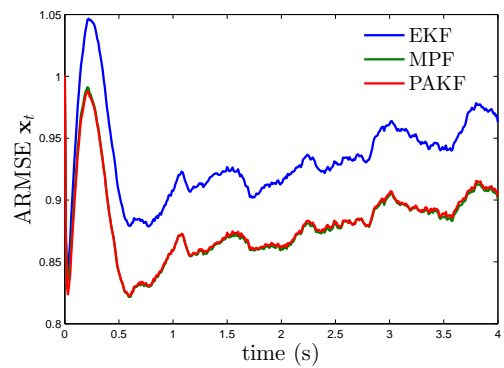}%
\caption{ARMSE for EKF, MPF and PAKF as a function of time.}%
%\caption{ARMSE as a function of time. The figure within shows a zoomed view  on a partition of the main plot.}%
\label{fig:rmse_mc_plot} %
\end{figure}

 Fig. \ref{fig:rmse_mc_plot}  shows the ARMSE between the simulated state and the estimated state as a function of time for the PAKF, MPF, and EKF.  We noticed that the ARMSE for EKF is always above the ARMSE of PAKF and MPF.
That is, PAKF and the MPF outperform EKF both for initial parts of the path and in the stabilized state. The MPF and  PAKF have similar performance, but MPF is computationally expensive.  
For 10\:000 particles, MPF takes six times more time to complete one MC run than PAKF. 

\section{Conclusion}
The proposed filter (PAKF) obtains estimation error close to that of the optimal filter (MPF) for a particular class of PWASS models which are discussed in this letter. The filter's performance is tested in  an example where the measurement noise variance is greater than the process noise variances and the comparison filters are EKF and MPF. The filtering recursion of PAKF involves approximation of the posterior distribution. Despite the approximations, PAKF obtains estimation error close to MPF and $5.35\%$ better than EKF. Furthermore, the computation time is roughly six times less than a MPF with comparable performance.

%
%=============================================================================
%===============================  New SECTION  ===============================
%=============================================================================
\clearpage

\bibliographystyle{IEEETran}
\bibliography{IEEEabrv,library}

%\appendix
 \begin{appendices}
\section{Mean and covariance matrix of DTMND}
\label{ap1}
A doubly-truncated multivariate normal distribution (DTMND) is a multivariate normal distribution, where one component is truncated from both below and above. 
Without loss of generality, we assume that the double truncation is applied to the first component of the random vector. For numerical methods, evaluating the presented formulas requires evaluation of the Cholesky decomposition \cite[Ch.\ 2.2.2]{bjorck_leastSquares} as well as the probability density function (PDF) and cumulative density function (CDF) of the univariate standard normal distribution.

\subsection{Formulas for mean and covariance matrix}

Let $\bx \in \reals^n$ be a random variable of the DTMND with the PDF
\begin{align}
p(\bx) \propto \N(\bx; \bmu, \cova) \cdot \ones_{[l_1,l_2]}([\bx]_1) ,
\end{align}
where $\bmu\in \reals^n$ is the location parameter vector, $\cova \in \reals^{n \times n}$ is the positive definite squared-scale matrix, and $l_1,l_2 \in \reals$ are the truncation limits. Further, let $\cholcov$ be the lower triangular matrix for which $\cova=\cholcov\cholcov^\t$ and whose diagonal entries are strictly positive. This type of square-root matrix can be obtained using the Cholesky decomposition \cite[Ch.\ 2.2.2]{bjorck_leastSquares}.

Then, the expectation value and covariance matrix of $\bx$ are
\begin{align}
\E[\bx] &= \Lambda \begin{bmatrix}  m^\ast  \\ \mathbf{0}_{n-1} \end{bmatrix}+ \bmu \label{eq:meanx} \\
\var[\bx] & = \Lambda \begin{bmatrix} s^\ast  & \mathbf{0}_{n-1}^\t \\ \mathbf{0}_{n-1} & I_{n-1} \end{bmatrix} \Lambda^\t
\label{eq:varx}
\end{align}
where
\begin{align}
m^\ast &=\frac{\phi(\lambda_1)-\phi(\lambda_2)}{\mathcal Z}, \label{eq:mast} \\
s^\ast &= 1+\frac{\lambda_1 \phi(\lambda_1)-\lambda_2\phi(\lambda_2)}{\mathcal Z} - (m^\ast)^2, \label{eq:sast}
\end{align}
with
\begin{equation*}
\lambda_1 = \frac{l_1-[\bmu]_1 }{[\Lambda]_{(1,1)}},\ 
\lambda_2 = \frac{l_2-[\bmu]_1 }{[\Lambda]_{(1,1)} },\
\mathcal Z = \Phi(\lambda_2)-\Phi(\lambda_1) .
\end{equation*}

\subsection{Derivation}

Let $\y \in \reals^n$ be a DTMND with the PDF
$$p_{\vct{y}}(\vct{y}) \propto \N(\vct{y}; \vct{0},I_n) \cdot \ones_{[\lambda_1,\lambda_2]}([\mathbf{y}]_1).$$
The components of $\mathbf{y}$ are independent, so the moments of $\mathbf{y}$ are obtained using the formula for the doubly-truncated univariate normal random variable \cite[Ch.\ 10.1]{johnson_ContUnivar1}. The mean and the covariance matrix are thus
\begin{align}
\E[\mathbf{y}] & =\begin{bmatrix} m^\ast \\ \vct{0}_{n-1} \end{bmatrix}, \label{eq:meany} \\
\var[\mathbf{y}] &=  \begin{bmatrix} s^\ast & \vct{0}_{n-1}^\t \\ \vct{0}_{n-1} & I_{n-1} \end{bmatrix}, \label{eq:vary}
\end{align} 
where $m^\ast$ and $s^\ast$ are those in \eqref{eq:mast} and \eqref{eq:sast}.

Let now $\vct{z} = \Lambda\vct{y}+ \bmu$. The PDF of $\vct{z}$ is then
\begin{equation}
p_\vct{z}(\vct{z}) = p_{\vct{y}}(\Lambda^{-1}(\vct{z}-\boldsymbol\mu)) \cdot \mathrm{det}\!\left(\tfrac{\ud \vct{y}}{\ud \vct{z}}\right).
\label{eq:pz}
\end{equation}
As $\Lambda$ is a lower triangular matrix, $[\Lambda^{-1}]_{(1,1:n)}=\begin{bmatrix}\tfrac{1}{[\Lambda]_{(1,1)}} & \vct{0}_{n-1}^\t\end{bmatrix}$, so\\\mbox{$[\mathbf{y}]_1 = ([\mathbf{z}]_1-[\bmu]_1)/[\Lambda]_{(1,1)}$.} Thus, \eqref{eq:pz} becomes
\begin{align}
p_\vct{z}(\vct{z}) &\propto \N(\cholcov^{-1}(\mathbf{z}-\boldsymbol \mu); \vct{0}, I) \cdot \mathrm{det}(\Lambda)^{-1}   \nonumber\\ 
&\hspace{2cm} \cdot  \mathbf{1}_{[\lambda_1,\lambda_2]}\left( \frac{[\mathbf{z}]_1-[\bmu]_1}{[\Lambda]_{(1,1)}} \right) \\
&= \N(\vct{z}; \boldsymbol \mu, \cova) \cdot \mathbf{1}_{\left[l_1, l_2 \right]}\left( [\mathbf{z}]_1 \right) ,
\end{align}
because $\cholcov\cholcov^\t=\cova$, $l_i = [\cholcov]_{(1,1)} \lambda_i + [\bmu]_1$ for $i\in\{1,2\}$ and $[\cholcov]_{(1,1)}$ is positive. That is, $\mathbf{z}$ has the same distribution as $\bx$, so the expected value and covariance matrix of $\bx$ are
\begin{align}
\E[\bx] &= \E[\mathbf{z}] = \cholcov \E[\mathbf{y}] + \bmu \label{eq:zmean} \\
\var[\bx] &= \var[\mathbf{z}] = \cholcov \var[\mathbf{y}] \cholcov^\t \label{eq:zcov} .
\end{align}
By substituting \eqref{eq:meany} and \eqref{eq:vary} to \eqref{eq:zmean} and \eqref{eq:zcov}, respectively, we get the formulas \eqref{eq:meanx} and \eqref{eq:varx}.
 \end{appendices}

\end{document}